\documentclass[aps,reprint,superscriptaddress,nofootinbib, notitlepage,prl]{revtex4-2}

\usepackage[utf8]{inputenc}
\usepackage{graphicx}
\usepackage{xcolor}
\usepackage{mathrsfs}
\usepackage{amsmath}
\usepackage{enumitem}
\usepackage[colorlinks=true,allcolors=blue]{hyperref}

\begin{document}

\title{Connecting anomalous elasticity and sub-Arrhenius structural dynamics in a cell-based model}

\author{Chengling Li}\email{chengling.li@emory.edu}
\affiliation{Department of Physics, Emory University, Atlanta, GA, USA}
\author{Matthias Merkel}
\affiliation{Aix Marseille Univ, Universit\'e de Toulon, CNRS, CPT (UMR 7332), Turing Centre for Living systems,
Marseille, France}
\author{Daniel M. Sussman}\email{daniel.m.sussman@emory.edu}
\affiliation{Department of Physics, Emory University, Atlanta, GA, USA}

\date{\today}

\begin{abstract}
	Understanding the structural dynamics of many-particle glassy systems remains a key challenge in statistical physics.
	Over the last decade, glassy dynamics has also been reported in biological tissues, but is far from being understood.
	It was recently shown that vertex models of dense biological tissue exhibit very atypical, sub-Arrhenius dynamics, and here we ask whether such atypical structural dynamics of vertex models are related to unusual elastic properties.
   It is known that at zero temperature these models have an elasticity controlled by their under-constrained or isostatic nature, but little is known about how their elasticity varies with temperature.
	To address this question we investigate the 2D Voronoi model and measure the temperature dependence of the intermediate-time plateau shear modulus and the bulk modulus. 
	We find that unlike in conventional glassformers, these moduli increase monotonically with temperature until the system fluidizes.
	We further show that the structural relaxation time can be quantitatively linked to the plateau shear modulus $G_p$, i.e.\ $G_p$ modulates the typical energy barrier scale for cell rearrangements.
	This suggests that the anomalous, structural dynamics of the 2D Voronoi model originates in its unusual elastic properties.
	Based on our results, we hypothesize that under-constrained systems might more generally give rise to a new class of ``ultra-strong'' glassformers.
\end{abstract}

\maketitle


The rheological behavior of disordered materials, in particular in the presence of thermal fluctuations, is still not fully understood.
In the athermal (zero-temperature) limit, it depends strongly on the Maxwell constraint-counting criterion \cite{Maxwell1864,Calladine1978,lubensky2015phonons}, which distinguishes three classes of systems.
Roughly, ``over-constrained'' systems have more constraints than degrees of freedom and are generally rigid.
Conversely, ``under-constrained'' systems have fewer constraints than degrees of freedom and are generally mechanically unstable.
Finally, ``isostatic'' systems have balanced constraints and degrees of freedom, and often behave as weak solids.

\begin{figure}[b]
	\centering
	\includegraphics[width=1\linewidth]{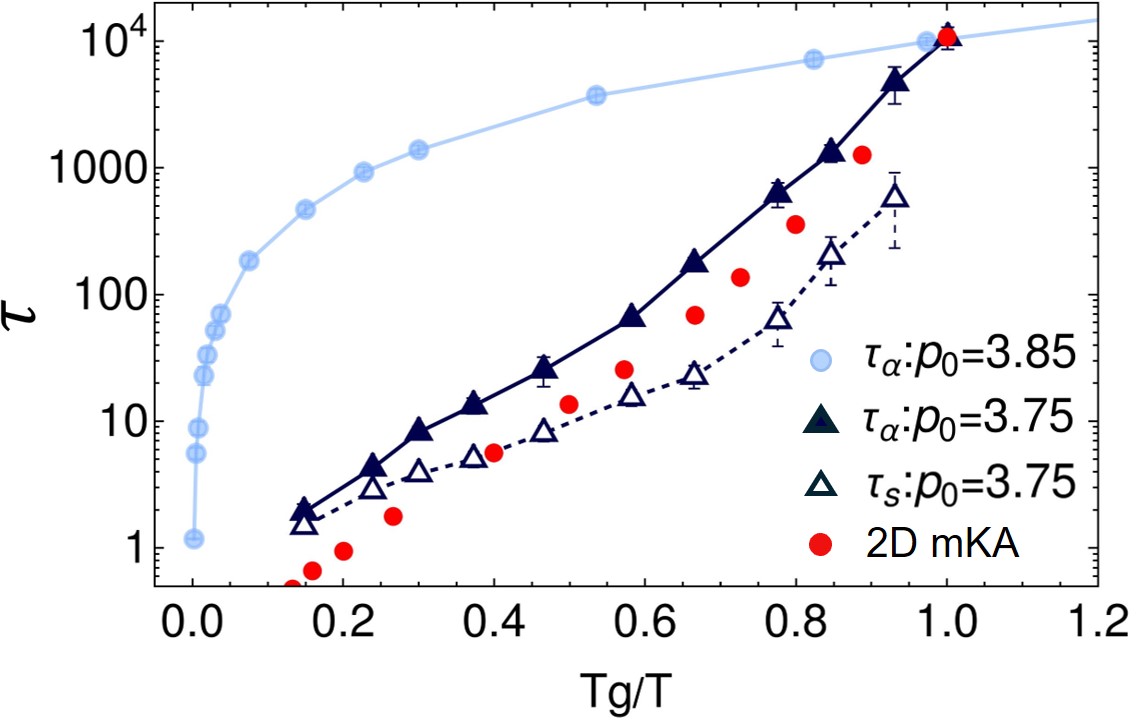}
	\caption{\textbf{Structural relaxation in the 2D Voronoi model}.
        Scaling of $\tau_\alpha$ of the 2D Voronoi model in the weak solid regime (light blue) and deeper in the solid regime (black), along with data from Ref. \cite{li2019long} for a standard 2D Kob-Andersen glassformer (red dots). Open symbols show the stress relaxation time, $\tau_s$.
	}
	\label{fig:tauAlphaFig}
\end{figure}

Over-constrained systems, which typically show glassy behavior, are the most well-understood.
Increasing temperature $T$ above the dynamical glass transition temperature $T_g$ facilitates particle rearrangements, which occur on a structural relaxation time scale $\tau_\alpha$.
A simple picture of structural relaxation based on temperature-independent energy barriers of order $\Delta E$ would predict that $\tau_\alpha\sim\exp{(\beta\Delta E)}$, where $\beta = (k_B T)^{-1}$ is the inverse temperature.
Such an exponential scaling of $\tau_\alpha$ with inverse temperature is called Arrhenius scaling, and standard particulate glasses display Arrhenius scaling or super-Arrhenius scaling (in which $\tau_\alpha$ grows faster than exponentially with the inverse temperature, c.f.\ the red dots in Fig.~\ref{fig:tauAlphaFig}) \cite{ediger1996supercooled}.
A long-standing idea is that since particle rearrangements require local elastic deformations of the material $\Delta E$, and thus $\tau_\alpha$, should be connected to elasticity \cite{cavagna2009supercooled,dyre2012instantaneous,puosi2012communication}.
Above $T_g$ at long times the shear relaxation modulus $G(t)$ vanishes and glasses are fluid, but this is often preceded by an intermediate-time plateau with value $G_p$.
Because transient deformations cost energy, the prediction of the elastic or ``shoving'' models is that $\Delta E\sim G_p$, i.e.\ \cite{dyre2012instantaneous,puosi2012communication}:
\begin{equation} 
    \tau_\alpha \sim \exp \left(\beta C G_p\right),\label{eq:shoving} 
\end{equation}
where $C$ is a constant.
In particulate glasses and other over-constrained systems $G_p$ typically varies extremely modestly with temperature, which would be consistent with Arrhenius behavior.
Super-Arrhenius behavior can arise form a slightly larger temperature dependence of $G_p(T)$ which can originate from subtle changes to the inherent states sampled as $T$ changes.

Under-constrained systems are less well understood.
The most commonly studied under-constrained systems have fixed connectivity, such as spring or fiber networks \cite{Merkel2019,chen2024field,broedersz2011criticality}.
While generally floppy at $T=0$, these systems can be rigidified through external strain or by varying model parameters in a way that introduces geometric incompatibility and prestresses.
These prestresses in turn govern elastic properties such as the shear modulus.
There is relatively little work on the thermal behavior of such systems.
A few specific networks have been studied, where the shear modulus was found to scale as a power law with temperature, $G\sim T^\alpha$, for $0.5\leq \alpha \leq 1$ \cite{mao2015mechanical,zhang2016finite}.
Recent analytical work by one of us on generic under-constrained networks found that for small temperatures there are three regimes; an energetically dominated regime with $\alpha=0$ close to the athermal rigid regime, an entropically dominated regime with $\alpha=1$ close to the athermal floppy regime, and a cross-over regime with $\alpha=0.5$ \cite{lee2023generic,lee2023partition}.
Thus, the shear modulus generally \emph{increases} with temperature.
However, because these models have fixed connectivity, no structural relaxation takes place.

There are also under-constrained models whose connectivity is not fixed, for instance so-called vertex models.
Vertex models describe dense biological tissues as tilings of polygonal or polyhedral cells  \cite{farhadifar2007influence,alt2017vertex,honda1983geometrical}.
They have been used to describe many experiments on biological tissues, including observations of cellular jamming transitions and glassy dynamics \cite{schoetz2013glassy,angelini2011glass,sadati2013collective,park2016collective,oswald2017jamming,park2015unjamming,garcia2015physics,tang2022collective,devany2021cell,grosser2021cell}.
Like under-constrained systems with fixed connectivity, athermal vertex models show a rigidity transition from a floppy to a rigid regime \cite{farhadifar2007influence,bi2015density,merkel2018geometrically,Merkel2019,duclut2021nonlinear,tong2022linear}.
There are also isostatic variants of vertex models, such as the 2D Voronoi model \cite{bi2016motility,sussman2018no,Pinto2022}.
There is no floppy regime in this model, but its shear modulus strongly decreases in the regime where the normal vertex model would becomes mechanically unstable \cite{sussman2018no}.
That is, the Voronoi model's behavior seems strongly affected by the athermal transition in under-constrained materials.
Importantly, in vertex and Voronoi models structural rearrangements occur as cells are allowed to change their local connectivity, and one of us showed that in the floppy or weak solid regimes, the structural dynamics of these models are of the atypical \emph{sub}-Arrhenius type (blue solid curve in Fig.~\ref{fig:tauAlphaFig}) \cite{sussman2018anomalous}.
Deep in the solid regime, on the other hand, they show a more normal super-Arrhenius behavior (black solid curve in Fig.~\ref{fig:tauAlphaFig}) \cite{li2021softness,Pandey2024}.

Are the atypical structural dynamics of vertex models related to the unusual elastic properties of under-constrained systems?
Due to its simplicity, we focus on addressing this question for the 2D Voronoi model.
We first measure its plateau shear modulus $G_p$ and its bulk modulus $K$. In the weakly solid regime, we find that both moduli increase with temperature, consistent with scaling exponents in the range $\alpha=0.5\dots1$.
In contrast, deep in the solid regime, we find that $G_p$ is independent of $T$ for $T\ll 1$ and decreases slightly for larger $T$.
We then compare this elastic behavior to the shoving model, Eq.~\eqref{eq:shoving}, and find data collapse when using a slightly modified form.
Our findings suggest that the anomalous structural dynamics of the Voronoi model originate in the unusual temperature dependence of its elastic properties.
This also explains why tuning the model from its weak to its deep solid regime leads to a change from sub- to super-Arrhenius structural dynamics \cite{li2021softness}.
Based on this, we expect to more generally observe sub-Arrhenius structural dynamics in under-constrained systems close to the athermal floppy regime.


We perform standard NVT simulations~\cite{martyna1996explicit,frenkel2002understanding} of the 2D Voronoi model under periodic boundary conditions using the open-source \emph{cellGPU} package~\cite{sussman2017cellgpu}.
In this model the positional degrees of freedom are the cell centers $\textbf{r}_i$; instantaneous Voronoi tessellations of $\{\textbf{r}_i\}$ determine the shapes of cells.
The many-body forces on the cells are defined as gradients of an energy functional with dimensionless form \cite{farhadifar2007influence}:
\begin{equation}\label{eq:cellenergyReduced} 
    E = \sum_{i=1}^{N}\Big[(a_i-1)^2 + (p_i-p_0)^2\Big].
\end{equation}
Here $N$ is the total number of cells, $a_i$ and $p_i$ are the area and perimeter of cell $i$.
The parameter $p_0$ corresponds to a preferred cell perimeter, and the preferred cell area, area modulus, and perimeter modulus have all been set to unity.
This energy is a biologically motivated \cite{honda1983geometrical,farhadifar2007influence} low-order expansion in the basic geometric properties of the cells \cite{kim2018universal}.
At $T=0$ the shear and bulk moduli of the Voronoi model are known to generally decrease with increasing $p_0$, especially in the weakly solid regime $p_0\approx3.8\dots3.87$ \cite{sussman2018no}.
We measure $\tau_\alpha$ by measuring the decay of the cage-relative self-intermediate scattering function \cite{vivek2017long, illing2017mermin}, and ensure that all simulations are at least $10$ times longer than the measured $\tau_\alpha$.
We report results for systems of $N=4096$ cells, and have checked that increasing this to $N=32768$ does not change our results.

\begin{figure}[h]
	\centering
	\includegraphics[width=1\linewidth]{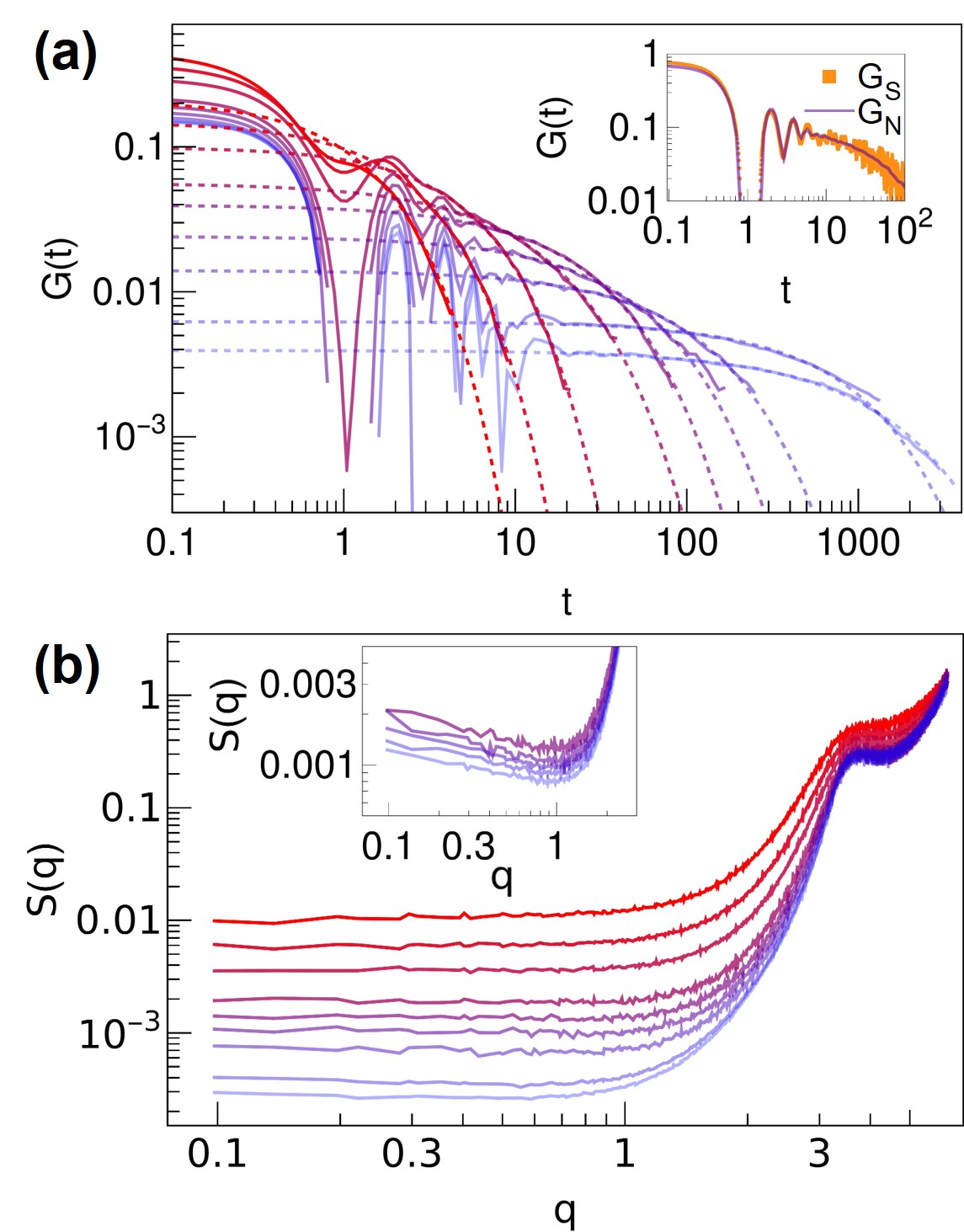}
	\caption{\textbf{Stress relaxation and static structure are strongly temperature-dependent}. 
		\textbf{(a)} Shear relaxation modulus as a function of time for $p_0=3.825$ and $T=\left[0.063 - 0.0005\right]$, decreasing from dark red to light blue.
		Dashed lines are stretched exponential fits.
		Inset: $G(T)$ at $T=0.0025$ computed via stress autocorrelations ($G_S$) or a small step strain ($G_N$).
		\textbf{(b)} The isotropic static structure factor $S(q)$ for $p_0=3.825$ and the same range of temperatures as in (a). 
		Inset: $S(q)$ for $p_0=3.8$ at $T=0.00385, 0.0031, 0.0028, 0.0025, 0.0022$.
	}
	\label{fig:GtSq}
\end{figure}

We first quantify the plateau shear modulus $G_p$.
At each simulation step we analytically calculate the shear stress $\sigma_{xy}\equiv(\sum_{i=1}^{N}d E_i/d \gamma)/A$, where $E_i$ is the energy of cell $i$, $\gamma$ is the shear strain, and $A$ is the total system area.
We then compute $G(t)$ via the stress auto-correlation function 
\begin{equation}\label{eq:G(t)}
	G(t)=\frac{A}{k_BT}\Big\langle\sigma_{xy}(t_0)\sigma_{xy}(t_0+t)\Big\rangle,
\end{equation}
where $\langle...\rangle$ represents an average over $t_0$ for each simulation, with statistics accumulated using a multiple-tau correlator method \cite{ramirez2010efficient}.
Figure~\ref{fig:GtSq}(a) shows sample $G(t)$ across a range of $T$ for $p_0=3.825$.
We verified these results by independently measuring $G(t)$ from the response to a small shear step $\Delta \gamma$ at $t=0$ via $G(t)=\Delta \sigma_{xy}(t)/\Delta\gamma$ (Fig.~\ref{fig:GtSq}(a) inset).
We find that the magnitude of $G(t)$ is strongly temperature-dependent at almost all time scales (Fig.~\ref{fig:GtSq}(a)).
As $T$ decreases, similar to other glassy systems \cite{flenner2019viscoelastic,puosi2012communication}, an intermediate-time plateau gradually emerges, followed by a final decay.
We quantify $G_p$ and the duration $\tau_s$ of the intermediate-time plateau by fitting our $G(t)$ data to a stretched exponential form, $G(t)\approx G_p e^{-(t/\tau_s)^\beta}$ at intermediate and late times (dashed curves in Fig.~\ref{fig:GtSq}(a)).
More precisely, we fit data from $10$ simulations for $t\in \left[t_i,t_f\right]$ for each set of parameters $(p_0,\ T)$; $t_i$ is chosen as the earliest time that an approximate plateau appears (and if there is no such plateau, we vary $t_i$ across a range to ensure a qualitatively good fit encompassing as many data points as possible), and $t_f$ is chosen as the time at which the magnitude of $G(t)$ is comparable to either the thermal fluctuations or the statistical error of our autocorrelation method \cite{ramirez2010efficient}.
We find that $\tau_s$ is typically several times smaller than $\tau_\alpha$, suggesting that in this model stress-bearing structures can dissipate before individual cells move a substantial amount.

\begin{figure}
	\centering
	\includegraphics[width=1\linewidth]{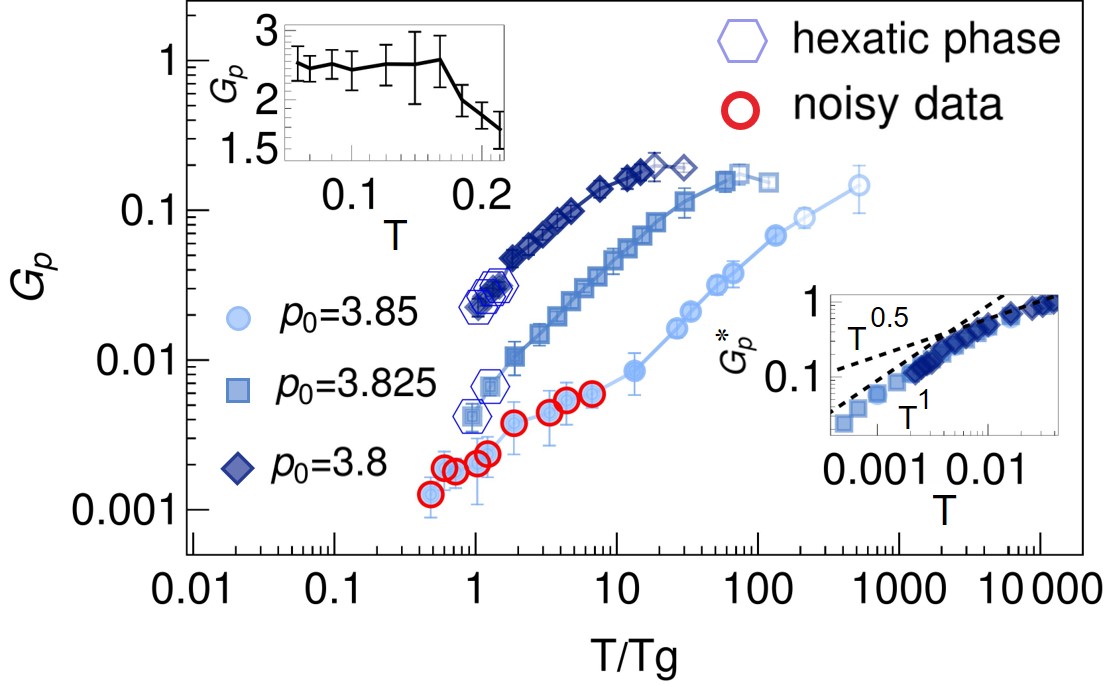}
	\caption{\textbf{The plateau shear modulus $G_p$ is strongly temperature-dependent.}
		Here state points corresponding to the hexatic phase are plotted by shown with a hexagon.
		At the lowest temperatures for $p_0=3.85$, the tail of $G(t)$ is both noisy and poorly fit by a stretched exponential; we consider these measurements less reliable and note them with red open circles.
		Lower Inset: $G_p^*=G_p/G_p(T=0.039)$ for each $p_0$ as a function of $T$.
		Dashed lines are guides to the eye of slope 1 and $1/2$.
        Upper Inset: temperature dependence of a $p_0=3.0$ Voronoi model whose structural relaxation is super-Arrhenius \cite{li2021softness}.
    }
	\label{fig:gpT}
\end{figure}

In Fig.~\ref{fig:gpT}, we plot the plateau shear modulus $G_p$ for $p_0=(3.8,3.825,3.85)$ as a function of $T/T_g$, with $T_g$ defined by $\tau_\alpha(T_g)\equiv 10^4$.
In contrast to particulate glassformers \cite{lu2009correlation,flenner2019viscoelastic,puosi2012communication}, in the Voronoi model $G_p$ strongly \emph{increases} upon increasing $T$ until at high enough temperatures we find a pure exponential decay of $G(t)$, which we associate with a fluid phase (Fig.~\ref{fig:gpT}).
At low temperatures the system enters a hexatic phase (hexagons in Fig.~\ref{fig:gpT}), as measured by a peak in the susceptibility of the bond-orientational order parameter \cite{halperin1978theory} -- this is consistent with earlier hexatic-phase observations in vertex models \cite{li2018role}.
Finally, we note that the $G(t)$ data for $p_0=3.85$ at the lowest temperatures was both noisy and poorly fit by a stretched exponential (red dots in Fig.~\ref{fig:gpT}), as we begin to hit the floor associated with the statistical error of our autocorrelator.
We have ignored this data in our subsequent analysis.

As shown in Fig.~\ref{fig:gpT}, $G_p$ shows a roughly power-law scaling of $T$, with $G_p\sim T^\alpha$.
In the lower inset, we rescale $G_p$ by its value in our highest-$T$ simulations.
We find excellent collapse of our data, with the scaling of $G_p(T)$ varying from $T^{0.5}$ to $T^1$.
Although no analytical calculations have been made for the Voronoi model, our results are consistent with observations in various fixed-connectivity spring-network models \cite{zhang2016finite,mao2015mechanical,arzash2023mechanical} and those predicted in generic under-constrained systems \cite{lee2023generic,lee2023partition}.

Given that the Voronoi model shows super-Arrhenius behavior deeper in the solid regime (black solid curve in Fig.~\ref{fig:tauAlphaFig}), we also measured the plateau shear modulus for the $p_0=3.0$ model studied in \cite{li2021softness} (Fig.~\ref{fig:gpT}, upper inset).
We find that $G_p$ is roughly constant at small temperatures and decreases for larger temperatures. That is, in the regime in which the model's dynamics mirror those of a more typical glassformer, so too does its elasticity.

\begin{figure}
	\centering
	\includegraphics[width=1\linewidth]{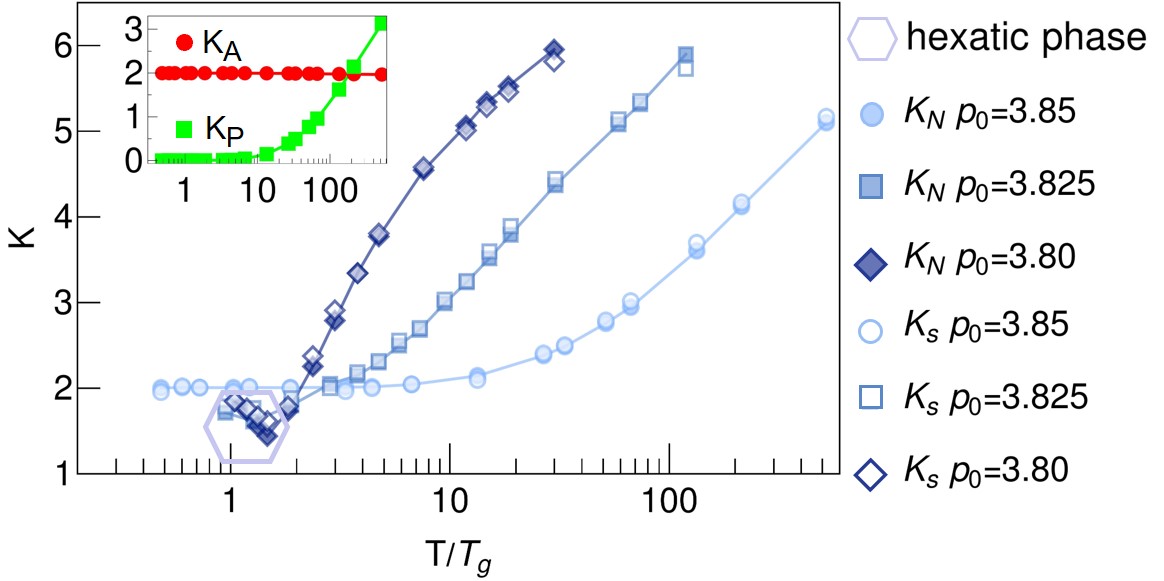}
	\caption{\textbf{The bulk modulus $K$ is strongly temperature-dependent.}
		The bulk modulus as calculated by $K_s=nk_BT/S(q_{min})$ (open symbols) and $K_N=dP/dA$ (closed symbols).
		Inset: Contributions to $K$ from area ($K_a$) and perimeter ($K_p$) terms in the energy.
	}
	\label{fig:K}
\end{figure}

We also characterize the temperature dependence of the bulk modulus $K$ by measuring the low-wavevector behavior of the static structure factor, $S(\textbf{q}) \equiv \frac{1}{N}\sum_{j=1}^{N}\sum_{k=1}^{N} e^{- i\textbf{q}\cdot (\textbf{r}_j-\textbf{r}_k)}$, where $j,k$ index the cells.
The bulk modulus is $K_s=Nk_BT/AS(q\rightarrow 0)$ \cite{hansen2013theory,zhuravlyov2023finite}.
Fig.~\ref{fig:GtSq}(b) shows representative plots of $S(q)$ across a range of $T$ for $p_0=3.825$.
For most of our systems there is an unambiguous low-$q$ plateau, although for the lowest temperatures for $p_0=3.8$ there is a non-monotonicity at small $q$ (Fig. \ref{fig:GtSq}(b)).
We thus use the minimum value attained by the structure factor, $S(q_\mathrm{min})$, as a lower bound for $S(q\rightarrow 0)$ at those temperatures.
We verified these results by directly measuring $K$ from the response to an isotropic compression strain of $10^{-3}$, computing the bulk modulus as $K_N=-A dP/dA$, for $P=-(\sigma_{xx}+\sigma_{yy})/2$.

In Fig.~\ref{fig:K}, we plot the bulk moduli obtained through both methods, $K_s$ and $K_N$, as a function of $T/T_g$ for $p_0=(3.8,3.825,3.85)$.
Both measurements are consistent, although we note that $K_s$ is larger than $K_N$ for some low-$T$ state points, consistent with our comments above.
Again in contrast to typical glassy systems, we find that $K$ increases monotonically with $T$: as the system is heated it becomes more resistant to isotropic deformations.
This temperature dependence of $K$ comes entirely from the perimeter term in Eq.~\eqref{eq:cellenergyReduced}. 
This is shown in the inset to Fig.~\ref{fig:K}, and is consistent with work on the athermal transition in vertex models, which was shown to be created entirely by the perimeter term \cite{sussman2018no,Merkel2019}.

\begin{figure}
	\centering
	\includegraphics[width=1\linewidth]{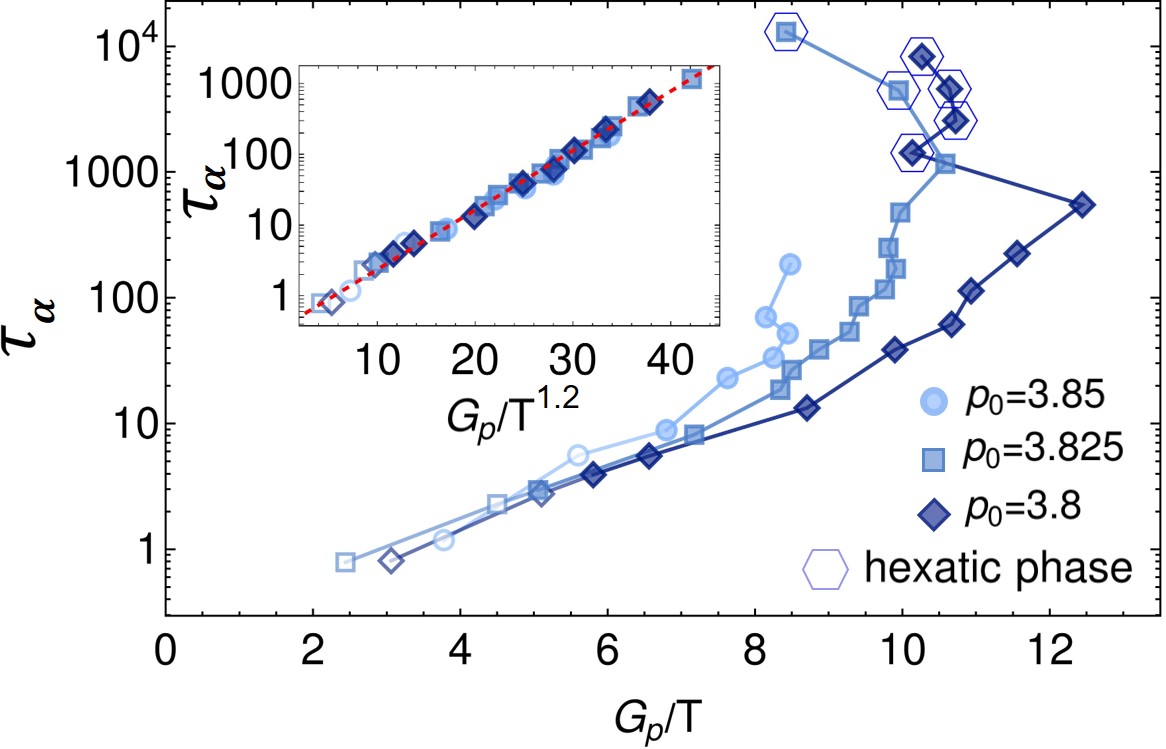}
	\caption{\textbf{The mechanics and dynamics are directly connected.}
		In the main plot we show $\log (\tau_\alpha) \ vs\ G_p/T$; the inset shows $\log (\tau_\alpha) \ vs\ G_p/T^{1.2}$.
		As in plots above, state points corresponding to hexatic phases or unreliable measurements of $G(t\rightarrow\infty)$ are noted. 
	}
	\label{fig:connection}
\end{figure}

Given the unusual temperature dependence of the Voronoi model's elastic moduli, does a shoving model such as Eq.~\eqref{eq:shoving} explain the anomalous structural dynamics?
In Fig.~\ref{fig:connection} we plot $\log(\tau_\alpha)$ vs. $G_p/T$ for the weak solid regime and find neither a linear scaling nor a compelling collapse of the data.
However, the inset shows excellent collapse when instead scaling $G_p$ by $T^{1.2}$.
This deviation from the typical elastic model suggests the presence of some other very modestly temperature-dependent feature contributes to the energy barrier $\Delta E$.


In summary, we have shown that the atypical sub-Arrhenius structural dynamics in the 2D Voronoi model is directly correlated to the unusual temperature dependence of its elastic moduli.
We specifically found that in the weak solid regime the plateau shear modulus scales with temperature as $G_p\sim T^{\alpha}$, with an exponent close to 1 for small temperatures and closer to 0.5 for larger temperatures, consistent with analytical results on fixed-connectivity under-constrained systems close to the athermal fluid regime \cite{lee2023generic,lee2023partition}.
Although the 2D Voronoi model is isostatic rather than under-constrained, we note the following.
First, at $T=0$ the elastic moduli of the Voronoi model have been shown to be governed by the rigidity transition of the vertex model \cite{sussman2018no}.
Second, in the deep solid regime of the 2D Voronoi model the structural dynamics becomes more conventional (black solid curve in Fig.~\ref{fig:tauAlphaFig}) \cite{li2021softness}, consistent with its constant plateau shear modulus in this regime (Fig.~\ref{fig:gpT}, upper inset); this in turn follows the prediction of a temperature-independent shear modulus in the solid regime of under-constrained systems \cite{lee2023generic,lee2023partition}.
Finally, in the 2D vertex model, the structural dynamics is known to be similar to that of the 2D Voronoi model \cite{sussman2018anomalous}, and the finite-temperature behavior of the elastic moduli close to the athermal fluid regime is expected to be extremely similar to what we report here for the Voronoi model \cite{lee2023generic,lee2023partition}.
Taken together, we expect that the reported sub-Arrhenius structural dynamics of the 2D vertex model is also explained by its elastic properties in the large-$p_0$ regime.
In the small-$p_0$ regime, in contrast, the more conventional structural dynamics \cite{Pandey2024} are likely related to the predicted more conventional elastic properties \cite{lee2023generic,lee2023partition}.

More generally, we would expect this scenario -- unusual elasticity connected to anomalous, sub-Arrhenius structural dynamics close to the athermal fluid regime, and more conventional elasticity and structural dynamics in other parts of parameter space -- to appear in any under-constrained system that is allowed to dynamically change its connectivity.
To our knowledge the only class of systems with glassy behavior similar to the vertex model are low-density vitrimeric polymers \cite{ciarella2019understanding}, where a network of covalently bonded under-constrained monomers can change its connectivity on long time scales via bond-exchange reactions.
We suggest that models with similar thermally dependent elasticity may all form a class of ``ultra-strong'' glasses, complementing the currently classification of ``fragile'' and ``strong'' glassformers.
If true, our findings may help to develop and tune materials with such an ultra-strong glass-forming ability.

Finally, our results have implications for the modeling of biological tissues.
Some modeling of dense tissue uses not shape-based under-constrained models but more typical particle-based models \cite{Germann2019,Henkes2020}.
Given the differences in elastic and structural properties between particulate glasses and vertex models, our work shows how this choice can lead to fundamentally different tissue-scale behavior.
For instance, our findings suggest that an increase in cellular motility would not only fluidify the tissue on long times \cite{bi2016motility}, but may also \emph{increase} the intermediate-time, tissue-scale rigidity.
It would be interesting to test this experimentally.

\begin{acknowledgments}
This material is based upon work supported by the National Science Foundation under Grant No.~DMR-2143815 (DMS and CL). 
This research used the Delta advanced computing and data resource which is supported by the National Science Foundation (award OAC 2005572) and the State of Illinois. Delta is a joint effort of the University of Illinois Urbana-Champaign and its National Center for Supercomputing Applications.
\end{acknowledgments}

\bibliography{finiteTemperatureMechanics}

\end{document}